\documentclass[10pt]{wlscirep}
\usepackage{bm,caption}

\newcommand{\gper}{\gamma_\perp}

\newcommand{\be}{\begin{equation}}
\newcommand{\ee}{\end{equation}}

\newcommand{\re}[1]{\mathop{\mathrm{Re}[#1]}}
\newcommand{\im}[1]{\mathop{\mathrm{Im}}[#1]}

\title{Nonlinear modal interactions in parity-time (${\cal PT}$) symmetric lasers}

\author[1,2,*]{Li Ge}
\author[3]{Ramy El-Ganainy}
\affil[1]{Department of Engineering Science and Physics, College of Staten Island, CUNY, Staten Island, NY 10314, USA}
\affil[2]{The Graduate Center, CUNY, New York, NY 10016, USA}
\affil[3]{Department of Physics and Henes Center for Quantum Phenomena, Michigan Technological University, Houghton, Michigan, 49931, USA}

\affil[*]{li.ge@csi.cuny.edu}

\begin{abstract}
Parity-time (${\cal PT}$) symmetric lasers have attracted considerable attention lately due to their promising applications and intriguing properties, such as free spectral range doubling and single-mode lasing. In this work we discuss nonlinear modal interactions in these laser systems under steady state conditions, and we demonstrate that several gain clamping scenarios can occur for lasing operation in the ${\cal PT}$-symmetric and ${\cal PT}$-broken phases. In particular, we show that, depending on the system’s design and the external pump profile, its operation in the nonlinear regime falls into two different categories: in one the system is frozen in the ${\cal PT}$ phase space as the applied gain increases, while in the other the system is pulled towards its exceptional point. These features are first illustrated by a coupled mode formalism and later verified by employing the Steady-state Ab-initio Laser Theory (SALT). Our findings shine light on the robustness of single-mode operation against saturation nonlinearity in ${\cal PT}$-symmetric lasers.
\end{abstract}

\begin{document}

\flushbottom
\maketitle
\thispagestyle{empty}

\section*{Introduction}

Motivated by fundamental studies in quantum systems \cite{Bender1,Bender2,Bender3}, the realization of ${\cal PT}$ symmetry in photonic structures have attracted considerable interest in the past few years \cite{El-Ganainy_OL07,Moiseyev,Musslimani_prl08,Makris_prl08,Guo,mostafazadeh,Longhi,CPALaser,conservation,Ge_PRX2014,Ruter,Lin,Feng,Feng2,Walk,Hodaei,Yang}. These structures are characterized by judiciously balanced gain and loss, and they exhibit a variety of intriguing light transport phenomena. The presence of gain has prompted several groups to study the concept of a ${\cal PT}$-symmetric laser \cite{Longhi,CPALaser}, which hosts several unique properties including free spectral range doubling as well as degenerate lasing and time-reversed lasing modes \cite{CPA,CPAexp}. Recently such lasers have been demonstrated using a micro-ring resonator with azimuthal complex index modulation \cite{Feng2} and two coupled micro-ring resonators \cite{Hodaei}, respectively. Both of them exhibit single-mode lasing behavior, which had not been anticipated before.

While linear threshold analysis (without considering nonlinearity) has revealed some important features of ${\cal PT}$-symmetric lasers \cite{Hodaei}, to which the single-mode lasing behavior was attributed, the laser is an intrinsically nonlinear system due to gain saturation, without which the system would not be stable. Therefore, it is important to consider nonlinear modal interactions in the analysis of such novel lasers, which is the goal of the present paper.

\begin{figure}[t]
\includegraphics[clip,width=\linewidth,bb = 0 0 695 220]{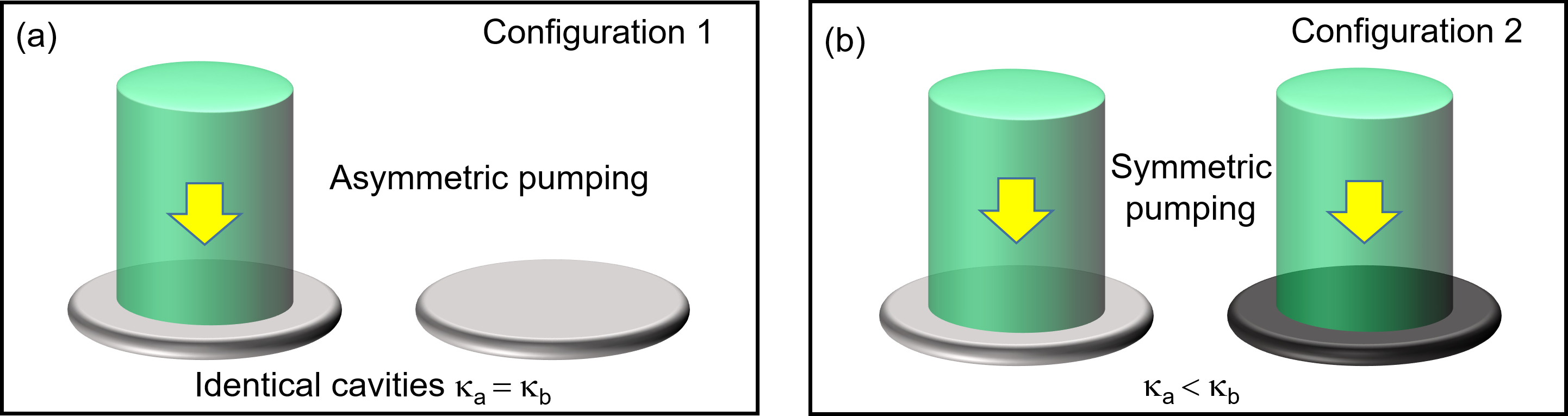}
\caption{Schematics of ${\cal PT}$-symmetric laser configurations discussed in the main text.
}\label{fig:schematics}
\end{figure}

We investigate two different ${\cal PT}$-symmetric laser configurations that represent essentially the setups in Refs.~\citenum{Feng2}
 and \citenum{Hodaei}.
This is first done by using a coupled mode formalism in Section ``Coupled mode analysis." We focus on a pair of supermodes that lie closest to the gain center $\omega_g$ that presumably lead to the lowest threshold.
In the first configuration [see Fig.~\ref{fig:schematics}(a)], we consider two identical cavities with the gain applied to only one of them (cavity $a$) \cite{Hodaei}. In this configuration a standard gain clamping behavior \cite{Lamb,Haken,Siegman} takes place once the laser is above its threshold, where the saturated gain maintains its threshold value \textit{independent} of whether lasing occurs in the ${\cal PT}$-symmetric phase or ${\cal PT}$-broken phase. As a result, the system is \textit{frozen} in the ${\cal PT}$ phase space, at a constant distance from its exceptional point (EP) \cite{EP1,EP2,EPMVB,EP3,EP4,EP5,EP6,EP_PRL,EP_exp,EP_CMT}, and the second supermode cannot reach its own threshold.
In the second configuration [see Fig.~\ref{fig:schematics}(b)], we consider equally applied gain to the two cavities, with the loss in cavity $b$ stronger than that in cavity $a$ (see the Discussion section for its connection with the setup in Ref.~\citenum{Feng2}).
Unlike the first configuration, here the gain clamping does not take place immediately above threshold if lasing occurs in the ${\cal PT}$-broken phase. Instead, the saturation effect takes place gradually as the applied gain increases. This gain saturation has a back action on the lasing mode and pulls the system towards its EP \cite{CREOL}.
While the modal gain of the second supermode is higher than its value in configuration 1, this mode is still suppressed even when the applied gain is high above its threshold value.

In Section ``SALT analysis," we examine these predictions using the Steady-state Ab-initio Laser Theory (SALT) \cite{Science,TS,TSG,SPASALT,C-SALT}, and we show that they hold qualitatively despite a weaker suppression of the second supermode. Furthermore, we extend the discussion of modal interactions by including other supermodes close to the gain center, which is beyond the scope of the simple coupled mode theory mentioned above.
This extension is important to determine the range of single-mode operation in ${\cal PT}$-symmetric lasers, and one key question is whether the
different gain clamping scenarios mentioned above can prevent all other supermodes from lasing, which would lead to an intrinsically single-mode laser.
While we found the answer to be negative, the modal interactions via gain saturation still lead to a wider range of single-mode operation (in terms of the applied gain) than previously expected from a linear threshold analysis.

\section*{Results}

\subsection*{Coupled mode analysis}

We first discuss modal interactions in ${\cal PT}$-symmetric lasers using a coupled mode formalism, where the gain saturation is incorporated under steady state conditions. The coupled mode approach is attractive due to its simple form that provides a physical insight into the role of coupling and non-Hermiticity (gain and loss) and how they affect the operation of ${\cal PT}$-symmetric lasers. In fact, this insight has broadened the definition of ${\cal PT}$-symmetric lasers to those without physically balanced gain and loss \cite{EP_CMT}, such as the ones considered in Refs.~\citenum{unconventional,EP_PRL,EP_exp}.

The coupled mode theory we employ takes the following form
\be
H =
\begin{bmatrix}
\omega_0 + i(\gamma_a-\kappa_a) & g \\
g & \omega_0 + i(\gamma_b-\kappa_b)
\end{bmatrix},\label{eq:H}
\ee
which acts on the wave functions $\varphi^{(\mu)}=[\psi_a^{(\mu)}~\psi_b^{(\mu)}]^T$ in cavity $a$ and $b$ (e.g., waveguides, microdisks, and microrings). Here ``$T$" denotes the matrix transpose. $\omega_{0}$ is the identical resonant frequency of the two cavities in the absence of coupling $g$, which is the closest one to the gain center $\omega_g$ and presumably corresponds to the lasing mode with the lowest threshold. $\kappa_{a,b},\gamma_{a,b}$ are the loss and saturated gain in the two cavities respectively, and we take $g$ to be a positive real quantity without loss of generality.

In configuration 1 mentioned in the introduction, we have $\kappa_a=\kappa_b$, $\gamma_b=0$, and $\gamma_a=\gamma/(1+\sum_\mu I^{(\mu)}_{a})$, where $\gamma$ is the applied gain and $I^{(\mu)}_{a}\equiv |\psi^{(\mu)}_{a}|^2$ is the intensity of mode $\mu$ in cavity $a$. $I^{(\mu)}_{a}$ is scaled by its natural units and dimensionless (see the discussion in Section ``SALT analysis"). This form of saturation is derived in steady state operation, with the fast dynamics of the polarization in the gain medium eliminated adiabatically.
In configuration 2, $\gamma_b$ is similarly defined [$\gamma_b=\gamma/(1+\sum_\mu I^{(\mu)}_{b})$] and nonzero, together with $\kappa_a<\kappa_b$. We note that the summations in $\gamma_{a,b}$ are only over the \textit{lasing} supermodes, i.e., the ones with a nonzero intensity.

To differentiate a lasing and non-lasing mode in our coupled mode theory, we note that the dynamics of the supermode $\mu$ here is given by $\varphi^{(\mu)}(t)=\varphi^{(\mu)}(0)\exp(-i\lambda^{(\mu)}t)$ in steady state operation, where $\lambda^{(\mu)}$ is one of the two eigenvalues of the effective Hamiltonian:
\be
\lambda = \omega_0 + i(\overline{\gamma}-\overline{\kappa}) \pm i\sqrt{(\Delta-\delta)^2-g^2}. \label{eq:lambda}
\ee
Here $\overline{\kappa},\overline{\gamma}$ are the averages of the losses and saturated gains of the two cavities, and
$\Delta, \delta$ are their half differences, i.e., $\Delta = (\kappa_b-\kappa_a)/2$, $\delta = (\gamma_b-\gamma_a)/2$.
A non-lasing mode does not exhibit a sustained laser oscillation with a finite amplitude, which indicates that the corresponding $\lambda$ has a negative imaginary part. A lasing mode, in contrast, features a real $\lambda$ in steady state that gives the lasing frequency.
The lasing threshold $\gamma_\text{TH}^{(\mu)}$ of mode $\mu$ can then be defined as the value of the applied gain $\gamma$ at which the corresponding $\lambda$ becomes real. For convenience, we will refer to $\im{\lambda}$ as the modal gain, which is negative for a mode below its threshold and becomes zero at and above its threshold.

Our coupled mode theory allows single-mode and two-mode operations, where one or both $\lambda$ given by Equation~(\ref{eq:lambda}) are real.
From the nonlinear optics point of view, this constraint on $I^{(\mu)}_{a,b}$ for a given $\gamma$ is very different from other models that have been applied to study steady states in ${\cal PT}$-symmetric systems \cite{Graefe,Segev,Ramezani}, where one imposes the constraint directly on nonlinearity, e.g., with a fixed total intensity $I_a^{(\mu)}+I_b^{(\mu)}$.
The nonlinearity reflected by $\gamma_{a,b}$ here represents modal interactions through gain saturation, including self saturation in the single-mode case and cross saturation as well in the two-mode case.

It should be noted that the effective Hamiltonian given by Equation~(\ref{eq:H}) is ${\cal PT}$-symmetric without requiring physically balanced gain and loss, i.e., with a net gain cavity and a net loss cavity [$\gamma_a-\kappa_a=-(\gamma_b-\kappa_b)$]. Instead, this balance holds with respect to the average gain and loss: the non-Hermitian part of $H$ is $\pm i(\Delta-\delta)$ on the diagonal after pulling out the common factor $i(\overline{\gamma}-\overline{\kappa})$ \cite{EP_CMT}:
\be
H =
[\omega_0 + i(\overline\gamma-\overline\kappa)]\bm{1} +
\begin{bmatrix}
i(\Delta-\delta) & g \\
g &  -i(\Delta-\delta)
\end{bmatrix},\label{eq:H2}
\ee
where $\bm{1}$ is the identity matrix.
Clearly it leads to an EP at
\be
|\Delta-\delta|=g.
\ee
Below we refer to the radicand in Equation~(\ref{eq:lambda}) as the ${\cal PT}$ parameter $\tau$:
\be
\tau \equiv (\Delta-\delta)^2-g^2.
\ee
The ${\cal PT}$-symmetric phase is defined by a negative $\tau$, where the modal gain of both supermodes are given by $(\overline{\gamma}-\overline{\kappa})$. The ${\cal PT}$-broken phase is defined by a positive $\tau$, the square root of which differentiates the modal gains of the two supermodes.

\subsubsection*{Configuration 1}

\begin{figure}[t]
\includegraphics[clip,width=\linewidth]{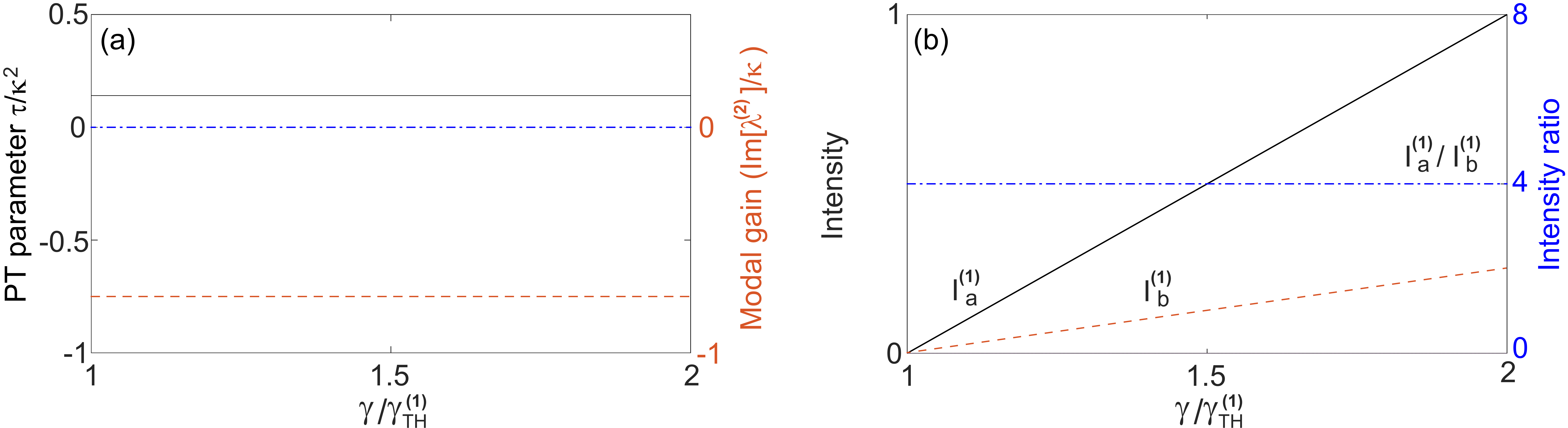}
\caption{Lasing in the ${\cal PT}$-broken phase in configuration 1a. (a) The frozen ${\cal PT}$ parameter $\tau=0.14\kappa^2$ (solid line) and the modal gain of the second mode $\im{\lambda^{(2)}}=-0.75\kappa$ (dashed line) as a function of the applied gain $\gamma$. The EP ($\tau=0$) and the lasing condition ($\im{\lambda}=0$) are marked by the dash-dotted line. (b) Intensity of the first mode in both cavity $a$ (solid line) and $b$ (dashed line). Their ratio is shown by the dash-dotted line, with its scale shown on the right side of the figure.
Here $\omega_0/\kappa=10^4$ and $g/\kappa=0.5$. 
}\label{fig:conf1a}
\end{figure}

We start with the discussion of nonlinear modal interactions in configuration 1, where $\Delta=0$ ($\kappa_a=\kappa_b\equiv\kappa$) and $\delta=-\gamma_a/2=-\overline{\gamma}$. We first investigate the ${\cal PT}$-broken phase (which we denote as case 1a), based on which single-mode lasing was demonstrated in Ref.~\citenum{Hodaei}.
This case requires $\kappa>g$ \cite{EP_CMT}, and the constraint of a real $\lambda$ becomes
\be
\overline{\gamma} = \overline{\kappa} \pm \sqrt{\delta^2-g^2}.
\ee
The ``$\pm$" signs represent the two supermodes, and it is easy to check that only the ``$-$" sign leads to a physical (real-valued) threshold given by
\be
{\gamma}^{(1)}_\text{TH} = \frac{\kappa^2+g^2}{\kappa} \label{eq:TH1a}
\ee
in terms of the applied gain $\gamma$.
To maintain a real $\lambda$ above threshold, it is straightforward to show that $\gamma_a=\gamma^{(1)}_\text{TH}$ must hold, i.e., the saturated gain (in cavity $a$) is clamped at its threshold value. Consequently, the system is frozen in the ${\cal PT}$-broken phase, with a constant
\be
\tau = \left(\frac{\kappa^2-g^2}{2\kappa}\right)^2>0
\ee
[see Fig.~\ref{fig:conf1a}(a)] and a constant intensity ratio above threshold:
\be
\frac{I^{(1)}_a}{I^{(1)}_b}
= \frac{\kappa^2}{g^2}>1
\ee
[see Fig.~\ref{fig:conf1a}(b)]. The intensity of mode 1 can be directly calculated from the clamped gain,
\be
I^{(1)}_a = \frac{\kappa}{\kappa^2+g^2}\gamma - 1,
\ee
and the modal gain of mode 2 stays negative and clamped [see Fig.~\ref{fig:conf1a}(a)], i.e.,
\be
\im{\lambda^{(2)}} = -2\sqrt{\tau}<0.
\ee
Therefore, the second mode is suppressed and cannot reach its threshold.

We note that the ${\cal PT}$-symmetric laser in this case does not have physical balance of gain and loss above threshold, because the net gain in cavity $a$ (given by $\gamma_a-\kappa=g^2/\kappa$) is smaller than the net loss in cavity $b$ (given by $\kappa$). This imbalance increases with $\tau$ and becomes significant deep in the ${\cal PT}$-broken phase. In contrast, lasing in the ${\cal PT}$-symmetric phase, defined by $\kappa<g$ and denoted by case 1b, does feature physically balanced gain and loss as we discuss below.

\begin{figure}
\includegraphics[clip,width=\linewidth]{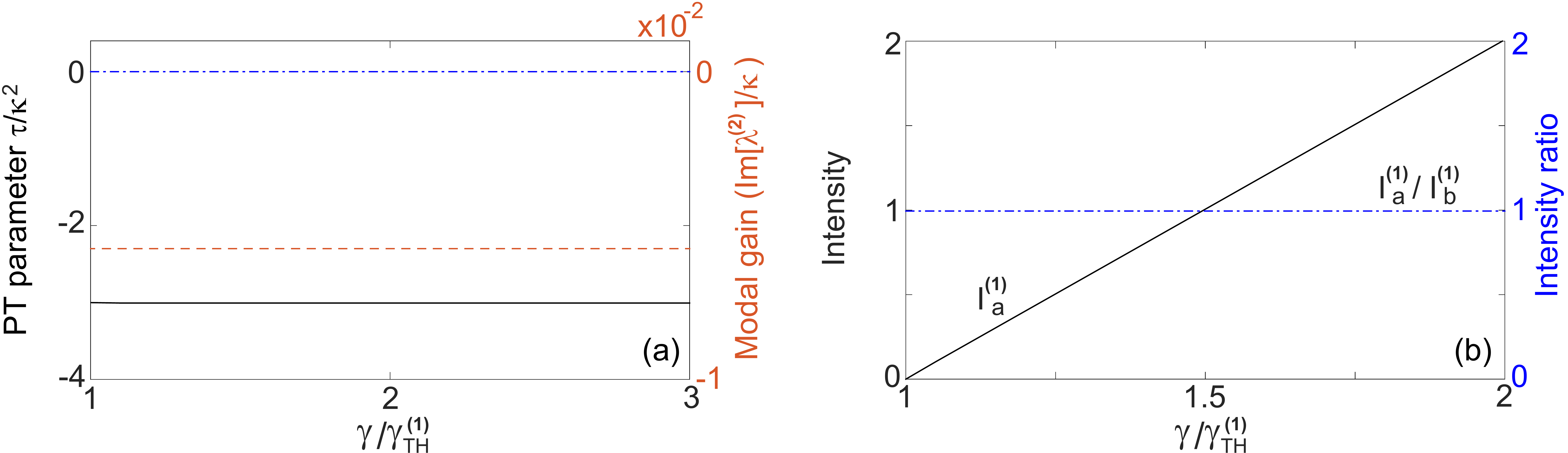}
\caption{Lasing in the ${\cal PT}$-symmetric phase in configuration 1b. (a) The frozen ${\cal PT}$ parameter $\tau=-3\kappa^2$ (solid line) and the modal gain of the second mode $\im{\lambda^{(2)}}<0$ (dashed line) as a function of the applied gain $\gamma$. The EP ($\tau=0$) and the lasing condition ($\im{\lambda}=0$) are marked by the dash-dotted line. (b) Intensity of the first mode in cavity $a$ (solid line). The ratio $I^{(1)}_a/I^{(1)}_b$ is shown by the dash-dotted line. Here $\omega_0/\kappa=10^4$ in cavity $a$ and slightly smaller (by $10^{-3}$) in cavity $b$. $g/\kappa=2$. 
}\label{fig:conf1b}
\end{figure}

In case 1b the ${\cal PT}$ parameter $\tau$ is negative and the modal gains of mode 1 and 2 are the same, given by $\gamma_a/2-\kappa$. Therefore, the constraint of a real $\lambda$ is given by
\be
\gamma_a = 2{\kappa},\label{eq:TH1b}
\ee
i.e., the saturated gain is clamped at its threshold value $2\kappa$, and above threshold the net gain in cavity $a$ (given by $\gamma_a-\kappa=\kappa$) equals the net loss in cavity $b$ (given by $\kappa$).
As a consequence of the gain clamping, the system is frozen in the ${\cal PT}$-symmetric phase, with a constant
\be
\tau=\kappa^2-g^2<0
\ee
[see Fig.~\ref{fig:conf1b}(a)].

These behaviors (i.e., gain clamping at threshold and a frozen ${\cal PT}$ parameter) are similar to those in case 1a, but the supermode symmetries here are different from those in case 1a. In particular, both supermodes here have a symmetric intensity profile $I^{(\mu)}_a=I^{(\mu)}_b$ and the same threshold. In reality only one of them lases, for example, due to a slight difference of the resonant frequencies in the two cavities. With this additional consideration and assuming mode 1 is the lasing mode, we find
\be
I^{(1)}_a = \frac{\gamma}{2\kappa} - 1
\ee
above its threshold [see Fig.~\ref{fig:conf1b}(b)], and the other supermode has a negative modal gain at the threshold of mode 1 [see Fig.~\ref{fig:conf1b}(a)]. Since the saturated gain is clamped, the modal gain of this mode is also clamped. As a result, this mode is suppressed and cannot reach its threshold.

As a final remark for configuration 1, we note that lasing in the ${\cal PT}$-broken phase (case 1a) is more favorable than lasing in the ${\cal PT}$-symmetric phase (case 1b): the threshold given by Equation~(\ref{eq:TH1a}) is lower than that given by Equation~(\ref{eq:TH1b}) for the same loss $\kappa$, which also leads to a stronger total intensity
\be
I^{(1)}_a+I^{(1)}_b =
\left\{
\begin{array}{ll}
\vspace{5pt}
\frac{\gamma}{\kappa} - (1+\frac{g^2}{\kappa^2}), & \text{case 1a}\,(\kappa>g),\\
\frac{\gamma}{\kappa} - 2, & \text{case 1b}\,(\kappa<g).
\end{array}
\right.
\ee
For evanescently coupled cavities, the coupling $g$ depends strongly on the inter-cavity distance $s$. Therefore, if $s$ is tuned and $\kappa-g$ changes sign as a result, one can imagine a transition between lasing in these two phases. For example, if the cavities undergo mechanically oscillations (``oscillating photonic molecule"), the laser output does not vary when the system stays in the ${\cal PT}$-symmetric phase, and it spikes periodically if $\max[s]$ is large enough to push the system into the ${\cal PT}$-broken phase.

\subsubsection*{Configuration 2}

In configuration 2 cavity $b$ has a higher loss than cavity $a$ ($\Delta>0$) and the gain is applied equally to both cavities. Note that the latter does not necessarily imply that $\delta=(\gamma_b-\gamma_a)/2$ is zero in the nonlinear regime, as we shall see below. The laser at threshold is ${\cal PT}$-symmetric (broken) if $\Delta<g$ ($\Delta>g$).

Lasing in the ${\cal PT}$-symmetric phase (case 2b) is similar to that in case 1b: the two supermodes have the same threshold now given by
\be
\gamma^{(1)}_\text{TH} = \overline\kappa,\label{eq:TH2b}
\ee
but in reality only one of them lases with equal intensities in the two cavities. Hence the applied gain $\gamma$ is saturated symmetrically ($\gamma_a=\gamma_b$ and $\delta=0$) as the applied gain increases, which indicates that the system is again frozen in the ${\cal PT}$-symmetric phase, with a constant ${\cal PT}$ parameter
\be
\tau=\Delta^2-g^2<0.
\ee
In addition, we find that $\gamma_{a,b}=\overline{\gamma}=\overline{\kappa}$ using the definition of $\overline{\gamma}$ and the constraint of a real $\lambda$, which shows that the saturated gains in both cavities are clamped at their threshold values. Thus mode 2 is prevented from lasing, with its modal gain staying below threshold. Meanwhile, we note that the laser features physically balanced gain and loss above threshold as in case 1a, because the net gain in cavity $a$ is given by $\gamma_a-\kappa_a=\Delta$ and equals the net loss in cavity $b$ (given by $\kappa_b-\gamma_b=\Delta$).
Finally, we find
\be
I^{(1)}_{a,b}=\frac{\;\gamma\;}{\overline{\kappa}}-1 \label{eq:int_symm}
\ee
above threshold using $\gamma_{a,b}=\overline{\kappa}$. All these behaviors are qualitatively the same as those shown in Fig.~\ref{fig:conf1b} and are hence not shown.



\begin{figure}[t]
\includegraphics[clip,width=\linewidth]{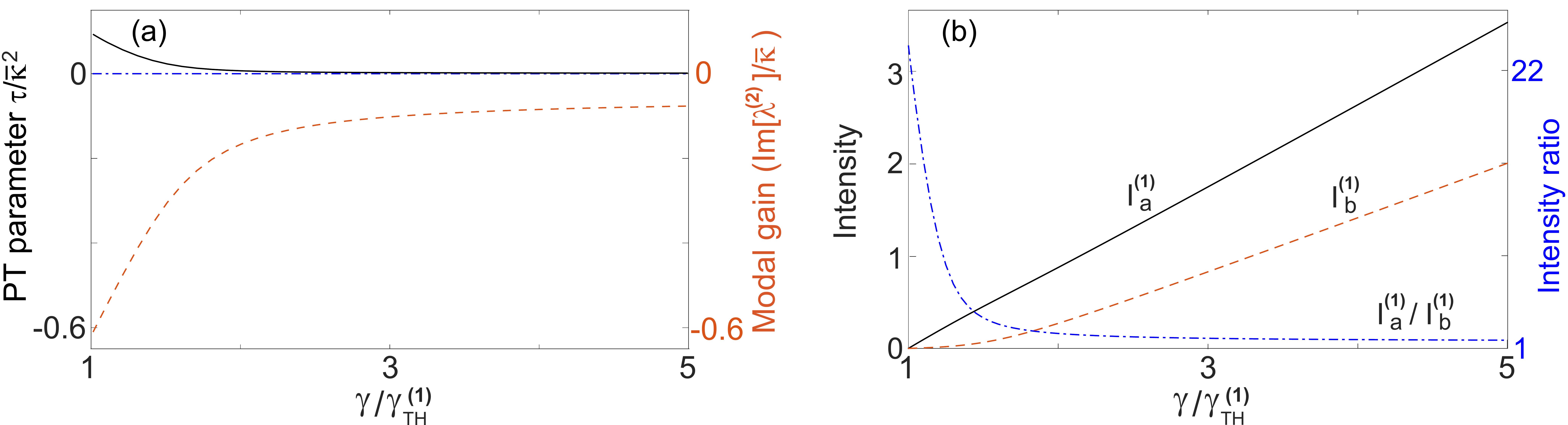}
\caption{Lasing in the ${\cal PT}$-broken phase in configuration 2a. (a) The ${\cal PT}$ parameter $\tau$ (solid line) and the saturated modal gain of the second mode $\im{\lambda^{(2)}}$ (dashed line) as a function of the applied gain $\gamma$. The EP ($\tau=0$) and the lasing condition ($\im{\lambda}=0$) are marked by the dash-dotted line. (b) Intensities of the first mode in both cavity $a$ (solid line) and $b$ (dashed line). Their ratio is shown by the dash-dotted line, with its scale shown on the right side of the figure.
Here $\omega_0/\kappa_a=10^4$, $\kappa_b/\kappa_a=2$ and $g/\kappa_a=0.2$. 
}\label{fig:conf2a}
\end{figure}

Lasing in the ${\cal PT}$-broken phase (case 2a) here is qualitatively different from the three cases (1a, 1b and 2b) discussed so far: the onset of the first lasing mode here does not lead to an immediate clamping of the gain as we show below. The constraint of a real $\lambda$ in this case is
\be
\overline{\gamma} = \overline{\kappa} \pm \sqrt{(\Delta-\delta)^2-g^2},\label{eq:constraint2}
\ee
and the laser threshold of the first supermode is given by
\be
\gamma_\text{TH}^{(1)}=\overline{\kappa} - \sqrt{\Delta^2-g^2},\label{eq:TH2a}
\ee
at which gain saturation just kicks in and $\delta=0$.
The intensity of the first mode is higher in cavity $a$ than in cavity $b$ above threshold:
\be
\frac{I^{(1)}_a}{I^{(1)}_b}
= \frac{(\sqrt{\tau}+\sqrt{\tau+g^2})^2}{g^2}>1.
\ee
Therefore, as the applied gain increases above $\gamma_\text{TH}^{(1)}$, it is saturated more in cavity $a$ than in cavity $b$, which leads to a positive and increasing $\delta$. As a result, the ${\cal PT}$ parameter $\tau=(\Delta-\delta)^2-g^2$ decreases towards zero [see Fig.~\ref{fig:conf2a}(a)].

In other words, this saturation has a back action on the lasing mode itself and the system is pulled towards its EP (where $\tau=0$) as a result: the intensity ratio $I^{(1)}_a/I^{(1)}_b$ reduces towards unity as $\gamma$ increases [see Fig.~\ref{fig:conf2a}(b)], or more precisely,
\be
\frac{I^{(1)}_a}{I^{(1)}_b}
\rightarrow \left[\frac{\overline{\kappa}+\Delta}{g+\sqrt{\overline{\kappa}^2+g^2-\Delta^2}}\right]^2,
\ee
the right hand side of which is approximate 1 when $\Delta\approx g$.
In addition, the saturated gains in both cavities approach their clamped values in the large $\gamma$ limit:
\begin{align}
\gamma_a &\rightarrow \kappa_a + \frac{g+\sqrt{\overline{\kappa}^2+g^2-\Delta^2}}{\overline{\kappa}+\Delta}g,\label{eq:gamma_a_2a}\\
\gamma_b &\rightarrow \kappa_b - \frac{\overline{\kappa}+\Delta}{g+\sqrt{\overline{\kappa}^2+g^2-\Delta^2}}g.\label{eq:gamma_b_2a}
\end{align}
This gain saturation then leads to an asymptotic value of the ${\cal PT}$ parameter:
\be
\tau = \frac{g^2}{4}\left[
\frac{g+\sqrt{\overline{\kappa}^2+g^2-\Delta^2}}{\overline{\kappa}+\Delta}
- \frac{\overline{\kappa}+\Delta}{g+\sqrt{\overline{\kappa}^2+g^2-\Delta^2}}
\right]^2\hspace{-5pt}>0.
\ee
In Fig.~\ref{fig:conf2a}(a) we have taken $g^2$ to be much smaller than $\overline{\kappa}^2$, and the above asymptotic value is very close to zero when measured by $\overline\kappa^2$.
We also note that the system does not have physically balanced gain and loss even in the large $\gamma$ limit: the net gain in cavity $a$ and the net loss in cavity $b$ are always reciprocal of each other, i.e.,
\be
(\gamma_a-\kappa_a)(\kappa_b-\gamma_b) = g^2;
\ee
they are equal only when the fractions in Equations~(\ref{eq:gamma_a_2a}) and (\ref{eq:gamma_b_2a}) become 1, or equivalently, $\Delta=g$.
Similar to case 1a, it's easy to show that the modal gain of mode 2 here is given by $\im{\lambda^{(2)}}=-2\sqrt{\tau}<0$ [see Fig.~\ref{fig:conf2a}(a)], meaning that mode 2 is also suppressed no matter how strong the applied gain is.

\subsection*{SALT analysis}

In the previous section we considered a pair of supermodes closest to the gain center $\omega_g$, one of which presumably is the first lasing mode when all the modes of the laser are considered. For other supermodes that are further away from the gain center, they typically have higher thresholds and lower modal gains. To understand the range of single-mode operation in ${\cal PT}$-symmetric lasers, it is important to take these additional supermodes into consideration. One key question we ask is whether the different gain clamping scenarios mentioned above can prevent other supermodes from lasing, which would lead to an intrinsically single-mode laser.

We probe this question using SALT \cite{Science,TS,TSG,SPASALT,C-SALT}, a semiclassical theory framework that addresses several key issues in the standard modal description of lasers \cite{Haken,Lamb} when applied to micro- and nano-systems. Most pertinent here is the inclusion of modal interactions to infinite order in SALT, without which artificial multimode lasing may appear shortly above the laser threshold \cite{OpEx08}.

\begin{figure}[t]
\includegraphics[clip,width=\linewidth]{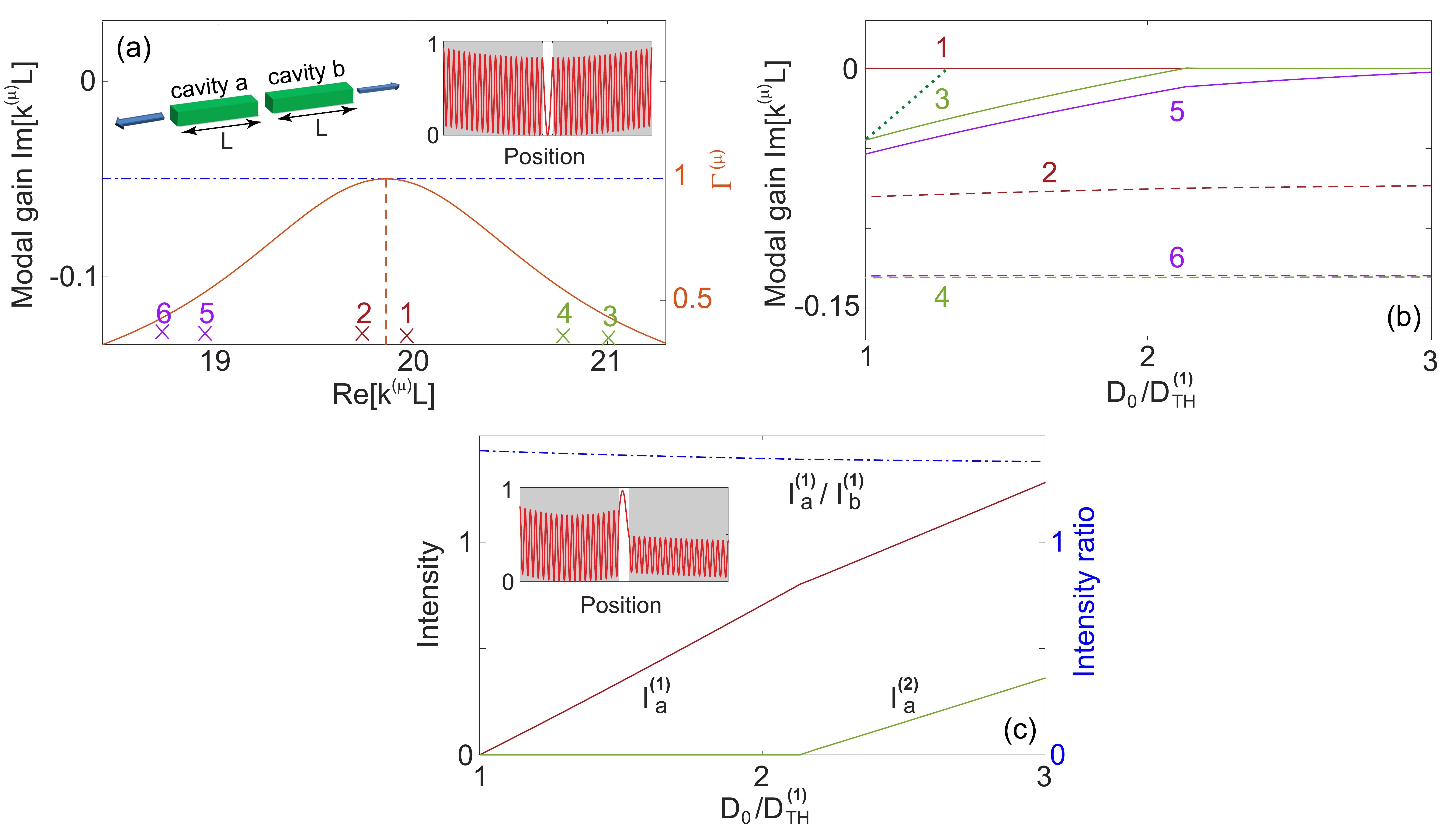}
\caption{A ${\cal PT}$-symmetric laser consists of two ridge cavities lasing in the ${\cal PT}$ broken phase. (a) 3 pairs of supermodes (crosses) underneath the gain curve (solid line). The vertical dashed line marks the gain center $\omega_g L/c=19.86$, and the dash-dotted line marks the lasing condition ($\im{k^{(\mu)}}=0$). Left inset: Schematics of the system. Right inset: Spatial profile of mode 1 before gain is applied. Grey areas indicate the two coupled cavities.
(b) and (c) Modal gain $\im{k^{(\mu)}}$ and intensity $I^{(\mu)}_{a}$ as a function of the atomic inversion $D_0$. Single-mode lasing operates until $D_0/D_\text{TH}^{(1)}\approx 2.1$. In (b) the dotted section shows the modal gain of mode 3 if its modal interaction with mode 1 is neglected. In (c) the intensity ratio $I^{(1)}_{a}/I^{(1)}_{b}$ is also shown, and the spatial profile of mode 1 at threshold is given by the inset. The cavity refractive index is $\sqrt{\epsilon_c}=3+0.007i$, and gap between the two cavities is $L/10$.
}\label{fig:conf1a_SALT}
\end{figure}

The first ${\cal PT}$-symmetric laser we consider consists of two coupled 1D ridge cavities [see Fig.~\ref{fig:conf1a_SALT}(a); left inset]. The background dielectric constant of the cavities is taken as $\epsilon_c=(3+0.007i)^2$, the imaginary part of which represents parasitic losses (material absorption, scattering loss, etc.) while the outcoupling loss is taken into consideration by an outgoing/radiation boundary condition \cite{TS}. The gain is applied only to cavity $a$, which has a center frequency $\omega_g L/c=19.84$ and a width of $\gper L/c=1$. Here $L$ is the length of one ridge cavity and $c$ is the speed of light in vacuum. This laser operates in the ${\cal PT}$-broken phase, which corresponds to configuration 1a discussed in the previous section.

We consider 6 supermodes closest to the gain center, each given by a quasi-bound (QB) mode of complex eigenvalue ${k}^{(\mu)}$ before the gain is applied [see Fig.~\ref{fig:conf1a_SALT}(a)], and the pair closest to the gain center (mode 1 and 2) have 19 intensity peaks in each cavity. The applied gain is increased via the atomic inversion $D_0$ and results in a total dielectric constant given by \cite{SPASALT}
\be
\epsilon(x) = \epsilon_c(x) + \frac{\gper D_0F(x)}{\re{{k}^{(\mu)}}c-\omega_g+i\gper},
\ee
where $F(x)$ is the spatial profile of the pump and has the value of 1 (0) in regions with (without) gain.
We note that $\epsilon(x)$, as defined above, is mode-dependent due to the different eigenvalues ${k}^{(\mu)}$ of the supermodes.

As soon as mode $\mu$ starts lasing, its QB eigenvalue ${k}^{(\mu)}$ becomes real and gives the lasing frequency (once multiplied by $c$).
Hence the modal gain here can be defined as $\im{k^{(\mu)}L}$, which is dimensionless and increases with $D_0$ in general before the first laser threshold $D_\text{TH}^{(1)}$.
The applied gain saturates above threshold with a spatial hole burning denominator $1+\sum_\mu\Gamma^{(\mu)}|\varphi^{(\mu)}(x)|^2$ \cite{TSG}, where $\Gamma^{(\mu)}=\gper^2/[\gper^2+(\re{{k}^{(\mu)}}c-\omega_g)^2]$ is the Lorentzian gain curve [see Fig.~\ref{fig:conf1a_SALT}(a)] and $\varphi^{(\mu)}(x)$ is the dimensionless magnitude of the electric field scaled by its natural units \cite{Science}.
We note that the summation over $\mu$ in the spatial hole burning denominator is again only over the lasing modes.

Although mode 1 has a symmetric intensity profile before the gain is applied [see Fig.~\ref{fig:conf1a_SALT}(a); right inset], the lack of gain in cavity $b$ leads the system to the ${\cal PT}$-broken phase, resulting in $I^{(1)}_a/I^{(1)}_b>1$ above threshold [see Fig.~\ref{fig:conf1a_SALT}(c); inset]. Here the intensities in the two cavities are defined by $I^{(\mu)}_{a,b}\equiv\int_\text{cavity a,b} |\varphi^{(\mu)}(x)|^2 dx/L$, which are also dimensionless.
In Fig.~\ref{fig:conf1a_SALT}(b) we see that the modal gain of mode 2 has a minute increase above the threshold of the first mode, which agrees qualitatively with the prediction of gain clamping given by the coupled mode theory shown in Fig.~\ref{fig:conf1a}(a).
To verify that the system is frozen in the ${\cal PT}$ phase space (i.e., with a fixed $\tau$ and intensity ratio $I^{(1)}_a/I^{(1)}_b$), we show $I^{(1)}_a/I^{(1)}_b$ in Fig.~\ref{fig:conf1a_SALT}(c): it barely changes from its value of 1.43 immediately above its threshold.

While these features agree well with the results of the coupled mode theory, the gain clamping does not hold for other supermodes, especially for mode 3 and 5 whose modal gains continue to increase above the first threshold with the applied gain. This behavior is common in microlasers \cite{TSG} and caused by non-uniform saturation of the gain: it is depleted more at the intensity peaks of mode 1, with ``holes" burnt in its spatial gain profile. Mode 3 and 5 have  different numbers of intensity peaks (20 and 18 in one cavity) from mode 1, hence they can utilize the increased gain where the intensity of mode 1 is weak. Nevertheless, their interactions with mode 1 still extend the range of single mode operation significantly: mode 3 would have started lasing at just $29\%$ above the first threshold without considering gain saturation [see dotted line in Fig.~\ref{fig:conf1a_SALT}(b)], while this fraction is in fact $110\%$ due to gain saturation.

\begin{figure}[t]
\includegraphics[clip,width=\linewidth]{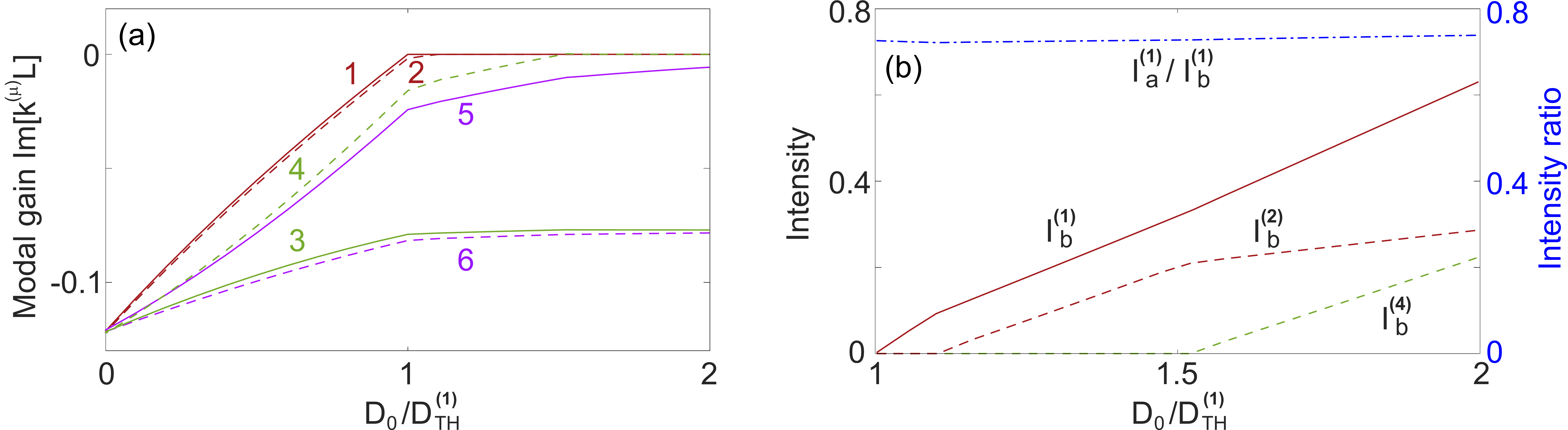}
\caption{A ${\cal PT}$-symmetric laser consists of two ridge cavities lasing in the ${\cal PT}$-symmetric phase. (a) and (b) show modal gain $\im{k^{(\mu)}}$ and intensity $I^{(\mu)}_{b}$ as a function of the atomic inversion $D_0$. In (b) the intensity ratio $I^{(1)}_{a}/I^{(1)}_{b}$ is also shown. The parameters are the same as in Fig.~\ref{fig:conf1a_SALT} except for $\epsilon_c=(3+0.001i)^2$, a gap width $L/40$, and $\omega_g L/c=20.03$.
}\label{fig:conf1b_SALT}
\end{figure}

Similar agreement with the coupled mode theory is observed in configuration 1b and 2b, and we show an example of the former in Fig.~\ref{fig:conf1b_SALT}. To make the system lase in the ${\cal PT}$--symmetric phase, we increase the coupling between the two cavities by shortening the gap between them (by a factor of 4) and reduce the cavity loss by having a smaller $\im{\sqrt{\epsilon_c}}=0.001$. We note that the difference of the modal gains for the supermode pair closest to the gain center is much smaller than in configuration 1a
[see Fig.~\ref{fig:conf1b_SALT}(a)], which indicates that the system is in the ${\cal PT}$-symmetric phase. The modal gain of mode 2 is still semi-clamped above threshold, but its minute increase beyond $D_\text{TH}^{(1)}$, similar to what we have seen in Fig.~\ref{fig:conf1a_SALT}(b), now pushes mode 2 above threshold shortly after mode 1 becomes lasing. This behavior eliminates configuration 1b (and 2b) as a candidate for single-mode operation. Another deviation from the result of the coupled mode theory lies in the spatial profile of mode 1 and 2: they are not necessarily symmetric above the first threshold, and in fact they have a similar intensity ratio $I_a/I_b<1$ in this example [see Fig.~\ref{fig:conf1b_SALT}(b)]. This is due to the outcoupling loss that is not considered in the coupled mode theory.
To be exact, the time reversal of a lasing mode at threshold is a coherent perfect absorption mode (``time reversed lasing mode") \cite{CPA,CPAexp} with purely incoming waves outside the system, hence a lasing mode itself does not satisfy ${{\cal PT}}\varphi^{(\mu)}(x)=\varphi^{(\mu)}(x)$ (which leads to $|\varphi^{(\mu)}(x)|^2=|\varphi^{(\mu)}(-x)|^2$ and $I^{(\mu)}_a=I^{(\mu)}_b$) even in the ${\cal PT}$-symmetric phase and with physically balanced gain and loss \cite{unconventional}. When the outcoupling/radiation loss is weak compared with the parasitic loss in a high-$Q$ cavity, for example, in coupled photonic crystal (PhC) defect cavities, we do recover $|\varphi^{(\mu)}(x)|^2\approx|\varphi^{(\mu)}(-x)|^2$ and $I^{(\mu)}_a\approx I^{(\mu)}_b$ at threshold (not shown).

\begin{figure}[t]
\includegraphics[clip,width=\linewidth]{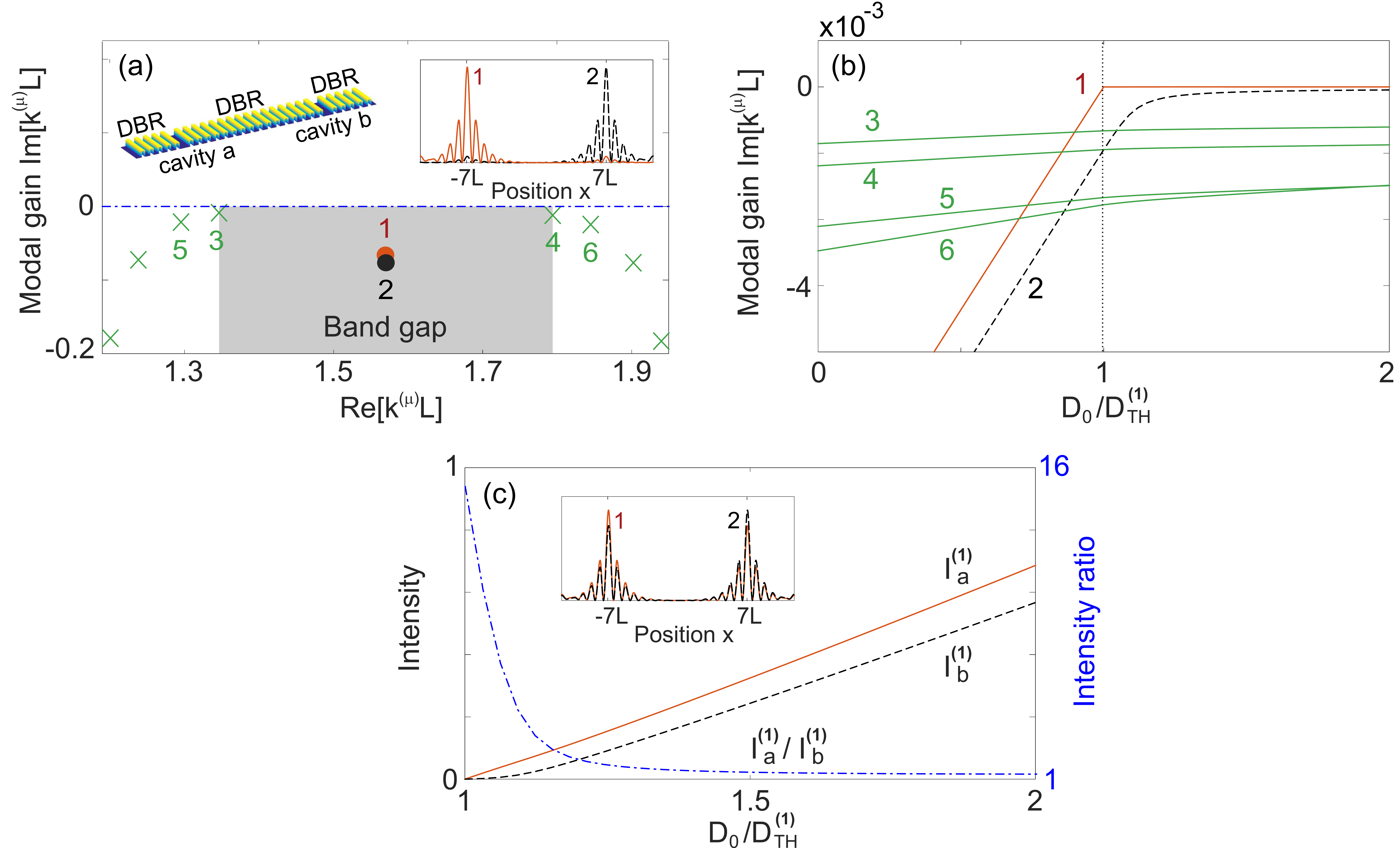}
\caption{A ${\cal PT}$-symmetric PhC laser in the ${\cal PT}$ broken phase. (a) Two defect modes in the band gap (circular dots) and several band edge modes (crosses). The dash-dotted line marks the lasing condition ($\im{k^{(\mu)}}=0$). Left inset: Schematics of the PhC laser. The DBRs consist of alternate layers with (1) index $2+0.001i$ and width $L/2$ and (2) index $3+0.001i$ and width $L/3$. $L$ is the length of the two cavities sandwiched by the DBRs. The refractive indices of the two cavities are $2+0.001i$ and $2+0.005i$. Right inset: Spatial profiles of mode 1 and 2 at the first threshold.
(b) and (c) Modal gain $\im{k^{(\mu)}}$ and intensity $I^{(\mu)}_{a,b}$ as a function of the atomic inversion $D_0$.
In (b) the vertical dotted line indicates the threshold.
In (c) the intensity ratio $I^{(1)}_{a}/I^{(1)}_{b}$ is also shown with its scale on the right side of the figure, and the inset shows the spatial profiles of mode 1 and 2 at the first threshold at $D_0=2D_\text{TH}^{(1)}$.
}\label{fig:conf2a_SALT}
\end{figure}

To exemplify lasing in configuration 2a, we use one-dimensional PhC defect cavities mentioned above. The whole system consists of two dielectric cavities $a$ and $b$ of length $L$ and refractive indices $n_c=2+0.001i,2+0.005i$, which are sandwiched by three distributed Bragg reflectors (DBRs) [see Fig.~\ref{fig:conf2a_SALT}(a); left inset].
We place the fundamental modes of the two cavities in the first band gap of the DBRs, and they couple to give rise to supermode 1, which is the lasing mode with an intensity ratio $I^{(1)}_a/I^{(1)}_b=15.1$ immediately above its threshold [see Fig.~\ref{fig:conf2a_SALT}(c)]. The other supermode 2 formed due to the coupling of these fundamental modes features $I^{(2)}_a/I^{(2)}_b=0.066\approx I^{(1)}_b/I^{(1)}_a$ at the same pump power, which indicates that lasing indeed occurs in the ${\cal PT}$-broken phase. The band edge modes in fact have a smaller loss ($|\im{k^{(\mu)}L}|$) before the gain is applied, but they are more extended and have a much weaker overlap with the applied gain in the two cavities. As a result, their modal gains increase much slower than those of the band gap modes [see Fig.~\ref{fig:conf2a_SALT}(b)], and they are suppressed even when the applied gain becomes very high. Meanwhile, the modal gain of mode 2 shows a clear saturation shortly above the threshold of the first mode and stays negative, which agrees well with the finding in the coupled mode theory. Finally, the system is pulled quickly towards its EP, which is manifested by the dramatic change of the intensity ratio $I_a/I_b$ of the first mode: it reduces quickly to 1.22 at $D_0=2D_\text{TH}^{(1)}$ [see Fig.~\ref{fig:conf2a_SALT}(c)].

We mention in passing that the same system can be used to demonstrate lasing in the ${\cal PT}$-symmetric phase (configuration 2b), if the gain is uniformly applied to both cavities and the DBRs. The band edge modes now become the lasing modes; their more extended spatial profiles lead to a stronger coupling between the two cavities, which overcomes the different losses of the cavities and leads the system to the ${\cal PT}$-symmetric phase. As a result, they have a more or less symmetric intensity profile with $I_a/I_b\approx1$.

\section*{Discussion}

\label{sec:conclusion}
In summary, we have discussed nonlinear modal interactions in the steady state of ${\cal PT}$-symmetric lasers, and we have shown different gain clamping scenarios that depend on (1) whether the loss or gain is uniform in the system and (2) whether lasing occurs in the ${\cal PT}$-symmetric or ${\cal PT}$-broken phase. As a consequence, the ${\cal PT}$-symmetric lasers can be separated into two categories: in one (I) the system is frozen in the ${\cal PT}$ phase space, while in the other (II) the system is pulled towards its exceptional point as the applied gain increases.

While the answer to the question imposed at the beginning of Section ``SALT analysis" (i.e., whether the ${\cal PT}$-symmetric lasers considered here are intrinsically single mode) is negative, 
the modal interactions via gain saturation in the ${\cal PT}$-broken phase do seem to indicate a robust single-mode operation even after all possible lasing modes are considered, and the range of applied gain for single-mode operation is significantly wider than previously expected from a linear threshold analysis \cite{Hodaei}. We emphasize that this behavior holds in high quality cavities too, as exemplified using a PhC defect laser, which is in contrast to the usual belief that ${\cal PT}$ symmetry related phenomena require considerable loss (and gain).

The two categories I and II cover many other gain and loss configurations that we haven't discussed. For example, if the two cavities have the same loss and the applied gains maintain a fixed ratio $\alpha\neq1$ in them (i.e., $\gamma$ in cavity $a$ and $\alpha\gamma$ in cavity $b$), lasing in its ${\cal PT}$-broken phase (which requires $\kappa^2/g^2>4\alpha/(\alpha-1)^2$) falls into category II with a weaker pulling effect, and lasing in its ${\cal PT}$-symmetric phase falls into category I. In all these cases, the lasing intensity is a monotonic function of the applied gain (at least in the single-mode regime), which is different from laser self-termination (LST) \cite{EP_PRL,EP_exp,EP_CMT} that requires a variable $\alpha$ (i.e., two independent pumps) as the applied gain increases. In fact, since the net gains (losses) in the two cavities discussed here also vary above threshold if the gain is not clamped immediately, we show below that LST can take place with a fixed $\alpha$ also (i.e., a single asymmetric pump) when the two cavities have different losses ($\kappa_b=\beta\kappa_a$).

\begin{figure}[h]
\includegraphics[clip,width=\linewidth]{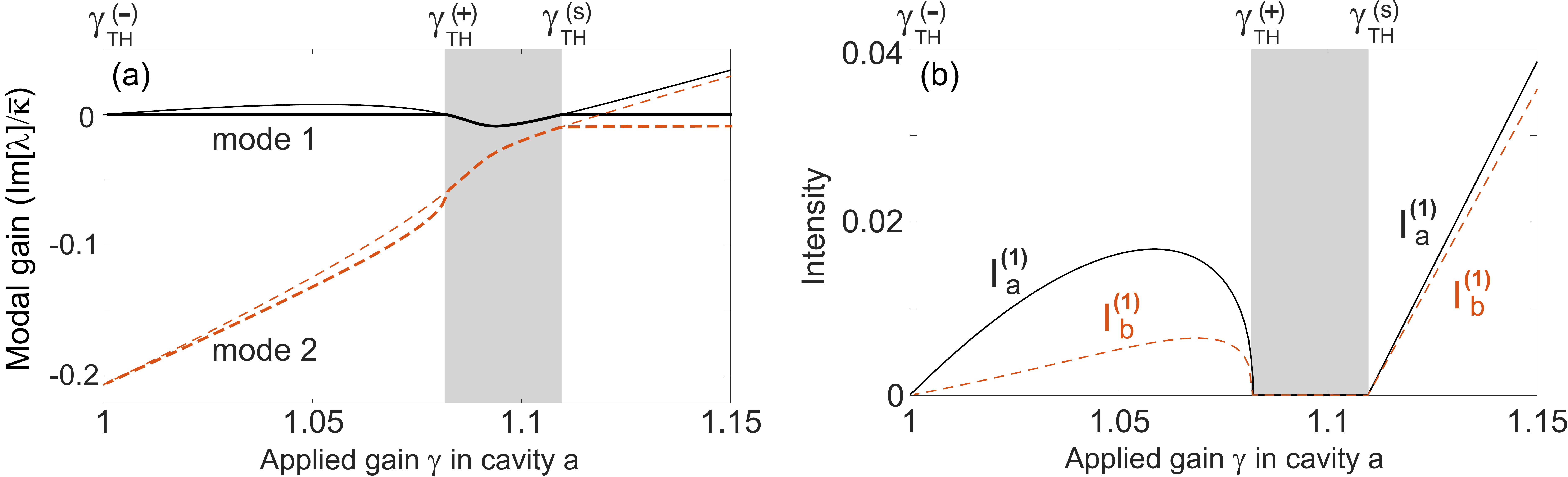}
\caption{Laser self termination in a ${\cal PT}$-symmetric laser with a single asymmetric pump beam. The applied gain $\gamma$ in cavity $a$ is plotted in unit of the first threshold $\gamma^{(-)}_\text{TH}$. The applied gain in cavity $b$ is $\alpha=3$ times as strong. (a) Modal gains (thick lines) of the two supermodes as a function of $\gamma$. Their values without gain saturations are given by the thin lines. The two sets of data coincide where the lasing mode is terminated (grey area). (b) Intensities $I_{a,b}^{(1)}$ of the lasing mode in the two cavities. Note that they are very different before the laser is terminated, indicating lasing in the ${\cal PT}$-broken phase, and they are very similar after the second onset threshold, indicating lasing in the ${\cal PT}$-symmetric phase.
Here $g/\kappa_a=1/3.6$, and $\kappa_b/\kappa_a=\beta=4$. $\omega_0/\kappa$ is $10^4$ in cavity $a$ and slightly lower (by $10^{-3}$) in cavity $b$. The slight detuning suppresses mode 2 beyond the threshold $\gamma_\text{TH}^{(s)}$ in the ${\cal PT}$-symmetric phase, similar to the situation in Fig.~\ref{fig:conf1b}. It also causes the slight imbalance between $I^{(1)}_a$ and $I^{(1)}_b$ beyond $\gamma_\text{TH}^{(s)}$. 
}\label{fig:LST}
\end{figure}

Our ${\cal PT}$-symmetric laser can have at most three thresholds, one in the ${\cal PT}$-symmetric phase given by $\overline{\gamma}=\overline{\kappa}$, or equivalently,
\be
\gamma_\text{TH}^{(s)} \equiv \frac{\beta+1}{\alpha+1}\kappa_a,\label{eq:TH_gen_symm}
\ee
and two in the ${\cal PT}$-broken phase given by $\overline{\gamma}=\overline{\kappa}\pm\sqrt{(\Delta-\delta)^2-g^2}$, or equivalently,
\be
\gamma_\text{TH}^{(\pm)} \equiv \frac{(\alpha+\beta)\kappa \pm \sqrt{(\alpha-\beta)^2\kappa^2-4\alpha g^2}}{2\alpha},\label{eq:TH_gen_broken}
\ee
where the radicand is positive. It is easy to check that we recover Equations~(\ref{eq:TH1a}) and (\ref{eq:TH2a}) from Equation~(\ref{eq:TH_gen_broken}) after taking $\beta=1,\alpha\rightarrow0$ and $\alpha=1$, respectively. Similarly, we recover Equations~(\ref{eq:TH1b}) and (\ref{eq:TH2b}) from Equation~(\ref{eq:TH_gen_symm}) in these two configurations. LST requires, in sequence, an onset threshold for mode 1 in the ${\cal PT}$-broken phase ($\gamma_\text{TH}^{(-)}$), a termination threshold for mode 1 in the same phase ($\gamma_\text{TH}^{(+)}$), and an onset threshold for mode 1 again in the ${\cal PT}$-symmetric phase ($\gamma_\text{TH}^{(s)}$) as the applied gain increases (see Fig.~\ref{fig:LST}(b) for example). In other words, LST requires $\gamma_\text{TH}^{(s)}$ given by Equation~(\ref{eq:TH_gen_symm}) to be higher than both $\gamma_\text{TH}^{(\pm)}$ given by Equation~(\ref{eq:TH_gen_broken}), which leads to
\vspace{10cm}
\be
\frac{4\alpha}{(\alpha-\beta)^2}<\frac{\kappa_a^2}{g^2}<\frac{4\alpha}{(\alpha-\beta)^2\left[1-(\frac{\alpha-1}{\alpha+1})^2\right]}
\ee
with the constraint $1<\alpha<\beta$ or $\beta<\alpha<1$. Following these criteria, we show one example of LST with $\alpha=3$, $\beta=4$, and $\kappa_a/g=3.6$ in Fig.~\ref{fig:LST} using the coupled mode theory.

Finally, we note that the single-mode lasing demonstrated in Ref.~\citenum{Feng2} is essentially configuration 2a we have discussed: the microring was patterned with strong and weak loss regions, and the gain is applied uniformly to the ring. Although in Ref.~\citenum{Feng2} the ${\cal PT}$ transition would be ``thresholdless" due to a Hermitican degeneracy, which was first reported by Ge and Stone in Ref.~\citenum{Ge_PRX2014} and later extended to a flat-band system \cite{flatbandPT}, the laser remains in the ${\cal PT}$-broken phase and does not differentiate whether the ${\cal PT}$ transition originates from an EP or a Hermitian degeneracy. We do note that the microring structure used in Ref.~\citenum{Feng2} does not allow lasing in the ${\cal PT}$-symmetric phase (configuration 2b), which does not actually exist when the non-Hermiticity of the system is nonzero.

\section*{Acknowledgements}

L.G. acknowledges support under PSC-CUNY Grant No. 68698-0046 and NSF Grant No. DMR-1506987. R.E. acknowledges support under NSF Grant No. ECCS-1545804.

\section*{Author contributions statement}

L.G. conceived the project, L.G. and R.E. performed the theoretical and numerical analyses.  Both authors wrote the manuscript.

\section*{Additional information}

The authors declare no competing financial interests.

\end{document}